\documentstyle[12pt,preprint,aps,prb,eqsecnum]{revtex}
\def\nm{\nonumber}
\def\beqa{\begin{eqnarray}}
\def\beq{\begin{equation}}   
\def\F{{\cal{F}}}

\def\eeqa{\end{eqnarray}}
\def\eeq{\end{equation}}
\def\lab{\label}
\def\pa{\partial}

\def\Del{\Delta}

\begin{document}

\begin{titlepage}
\thispagestyle{plain}
\pagenumbering{arabic}
\begin{center}
{\Large \bf Direct Derivation of Scaling Relation of Prepotential in}  
\end{center}
\vspace{-7.0mm}
\begin{center}
{\Large \bf $N=2$ Supersymmetric G$_2$ Yang-Mills Theory}
\end{center}
\vspace{-7.0mm}
\lineskip .80em
\vskip 4em
\normalsize
\begin{center}
{\large Y\H uji Ohta}
\end{center}
\vskip 1.5em
\begin{center}
{\em Research Institute for Mathematical Sciences }
\end{center}
\vspace{-11.0mm}
\begin{center}
{\em Kyoto University}
\end{center}
\vspace{-11.0mm}
\begin{center}
{\em Sakyoku, Kyoto 606, Japan.}
\end{center}
\begin{abstract}
In contrast with the classical gauge group cases, any method to prove 
exactly the scaling relation which relates moduli and prepotential is not 
known in the case of exceptional gauge groups. 
This paper provides a direct method to establish this relation 
by using Picard-Fuchs equations. In particular, it is 
shown that the scaling relation found by Ito in $N=2$ supersymmetric G$_2$ 
Yang-Mills theory actually holds exactly. 
\end{abstract}
\end{titlepage}
\pagestyle{myheadings}
\markboth{Derivation of scaling relation}
{Derivation of scaling relation}
\begin{center}
\section{Introduction}
\end{center}

It would be one of the greatest discoveries in 1990s that the holomorphic 
structure of the low energy effective prepotential of $N=2$ supersymmetric 
Yang-Mills theory in four dimensions was actually related with the moduli 
space of a Riemann surface which possesses singularities when charged 
particles become massless. According to this mechanism found by 
Seiberg and Witten, \cite{SW} the effective theory is parameterized by 
the vacuum expectation value of scalar components $\phi$ of 
$N=1$ chiral multiplet, which 
can be identified with periods of a certain meromorphic differential on the 
Riemann surface, and accordingly the prepotential can be determined exactly 
also including instanton corrections. For example, for classical gauge 
group cases, the prepotential is known to be dictated by hyperelliptic 
curves, \cite{AF,APS,KLT,DS1,BL,HO,Han,KLYT,AAG,MN} and instanton 
corrections to the prepotential obtained from these curves showed a good 
agreement with the so-called instanton 
calculus, \cite{FP,IS1,IS2,DKM,AHSW,HS,Slat,Yung} a pure field theoretical 
method. 

These hyperelliptic curves were also derived from a very different 
view point, relation to integrable systems. \cite{GKMMM,MW,IM} 
In the language of integrable system, 
these hyperelliptic curves coincide with the 
spectral curves (the characteristic equation for Lax matrix) and the 
periods can be interpreted as the action integrals. This interpretation 
explains why the effective theory is solvable, and from this fact, 
it may be natural to expect that also for exceptional 
gauge group cases the relevant curves are given by hyperelliptic curves. 
In fact, several hyperelliptic curves for such cases were 
constructed, \cite{AAG,DS2,AAM} but, unfortunately, in general, the 
spectral curves from integrable systems for exceptional gauge group cases can 
not be transformed 
into hyperelliptic form. For instance, the curve for G$_2$ gauge theory is 
related to the $(\mbox{G}_{2}^{(1)})^{\vee}$ Toda system, \cite{MW,LPG} 
and is given in the form   
	\beq
	3\left(z-\frac{\mu}{z}\right)^2 -x^8 +2ux^6 -\left[u^2 +
	6\left(z+\frac{\mu}{z}\right)\right]x^4 +\left[
	v+2u\left(z+\frac{\mu}{z}\right)\right]x^2 =0
	,\lab{12}
	\eeq
where $x$ is the eigenvalue of the Lax operator matrix, $z$ is the spectral 
parameter, $u$ and $v$ are 
gauge invariant Casimirs called moduli (of the effective theory) and $\mu$ 
is a parameter which leads to the dynamical scale, 
but obviously this curve can not be transformed into hyperelliptic 
form. Of course, for these two formulations, 
which one is better must be decided by a comparison of instanton 
corrections to prepotential with that from instanton calculus. According to 
the result, \cite{Ito} there is a manifest difference between prepotentials 
from these two curves and only that from (\ref{12}) can survive! 
Similar result follows also in E$_6$ gauge theory, \cite{Ghe} and 
therefore the substance of complex curves relevant to gauge theories is 
believed as a spectral curve of integrable system. 

As a characteristic common problem concerning to these gauge theories with 
exceptional Lie groups, we can mention that the exact establishment of 
scaling relation \cite{Mat,STY,EY,HW,KO} of prepotential $\F$ whose general 
form is typically represented in the form 
	\beq
	\sum_{i=1}^{g}a_i \frac{\pa \F}{\pa a_i} -2\F =\beta
	\langle \mbox{Tr}\,\phi^2 \rangle 
	\lab{fo}
	,\eeq
where $g$ is the rank of the gauge group, $a_i$ are the periods 
and $\beta$ is the coefficient of one-loop beta function, is very hard. 
Although the formula (\ref{fo}) has been checked in the standpoint of 
instanton calculus in the SU(2) gauge theory, \cite{FT,DKM2} 
as for a general proof of this formula for theories 
with any classical gauge groups, we know the method \cite{EY} based on 
Whitham hierarchy. \cite{NT} In such cases, the curves 
are hyperelliptic, and thus the verification of (\ref{fo}) can be done 
directly, but in contrast with these cases, for theories with exceptional 
gauge groups, similar discussions do not exist because of too complicated 
singularity structure of the spectral curves. For instance, in the case of 
the G$_2$ theory, we can explicitly see this from (\ref{12}), which is 
actually an eight cover of $z$-plane. 

On one hand, of course, there are strong supports for this formula 
(\ref{fo}) in exceptional gauge 
group cases and the validity of (\ref{fo}) was explicitly checked by 
first several instanton process levels with the help of explicit solutions 
to Picard-Fuchs equations, \cite{Ito,Ghe,GSA} but this does not 
mean that (\ref{fo}) holds exactly. Then, also in the theories with 
exceptional groups, does (\ref{fo}) hold exactly? To answer this question 
is the subject of this paper. 

In this paper, we give a method to verify (\ref{fo}) exactly for G$_2$ 
gauge theory, although the method itself is applicable for all theories 
with any classical and exceptional gauge groups with or without massive 
hypermultiplets. Our starting point is to 
consider the differentiated versions of (\ref{fo}), defined by
	\beq
	W=\sum_{i=1}^2 (a_i \pa_u a_{D_i}-a_{D_i} \pa_u a_i ),\quad 
	w=\sum_{i=1}^2 (a_i \pa_v a_{D_i}-a_{D_i} \pa_v a_i ) 
	\lab{sca}
	,\eeq
and seek differential equations for (\ref{sca}). As a matter of fact, this 
can be proceeded by considering Picard-Fuchs equations, but since the 
Picard-Fuchs equations with multiple moduli are usually realized as a set 
of partial differential equations, \cite{IMNS,Ali} actually such equations 
do not have any advantages for a study of scaling relation of prepotential 
(in higher rank gauge groups), although the derivation using partial 
differential 
form of Picard-Fuchs equations was tried in the case of SU(3). \cite{BM} 
However, if Picard-Fuchs equations represented by single kind of moduli 
derivatives can be found, we can easily construct an ordinary differential 
equation for respective quantities in (\ref{sca}). In addition, the basis 
of solutions to such ordinary differential equation can be uniquely fixed, 
so as the result, we can determine the right hand side of (\ref{fo}) 
by taking an appropriate ``initial condition''.

\begin{center}
\section{Picard-Fuchs equations in G$_2$ theory}
\end{center}

\begin{center}
\subsection{G$_2$ Picard-Fuchs equations}
\end{center}

To begin with, let us recall the Seiberg-Witten meromorphic differential 
on the G$_2$ curve (\ref{12}) given by 
	\beq
	\lambda =x\frac{dz}{z}
	\eeq
and the definition of periods 
	\beq
	a_i =\oint_{\alpha_i} \lambda ,\quad a_{D_i}=\oint_{\beta_i}
	\lambda ,\quad i=1,2
	,\eeq
where $\alpha_i$ and $\beta _i$ are the canonical cycles on (\ref{12}). 
Some of the properties of these period integrals were discussed by 
Masuda {\em et al.}. \cite{MSS} 
The classical relation among periods and moduli is given by \cite{AAM}
	\beq
	u=a_{2}^2 +(a_1 -a_2 )^2 +(a_1 -2a_2 )^2 ,\quad 
	v=a_{2}^2 (a_1 -a_2 )^2 (a_1 -2a_2 )^2 
	.\lab{lukeskywalker}
	\eeq
These are invariant under the action of the Weyl group of G$_2$ 
	\beq
	(a_1 ,a_2 ) \rightarrow (3a_2 -a_1 , a_2), \quad 
	(a_1 ,a_2 )\rightarrow (a_1 ,a_1 -a_2 )
	.\eeq

Then, the Picard-Fuchs equations can be given in the form \cite{Ito} 
	\[\Biggl[\frac{2(720u^2 \mu +2u^3 v-27v^2)}{-uv +24\mu}\pa_{u}^2 +
	\frac{4(256u^4 \mu -3u^2 v^2 -720uv\mu +13824\mu^2 )}{-uv +24\mu}
	\pa_u \pa_v \]
	\vspace{-1.0cm}
	\[-\frac{6(-256u^3 \mu +96 v\mu +5uv^2 )}{-uv +24\mu}\pa_v -1
	\Biggr]\lambda =0,\]
	\vspace{-1.0cm}
	\beq
	\Biggl[\frac{1}{3}(8u^3 v -108v^2 +2880u^2 \mu)\pa_{v}^2 
	+\frac{1}{3}(8u^4 -72uv +6912\mu )\pa_u \pa_v +(4u^3 -24v )\pa_v -1
	\Biggr]\lambda=0
	.\lab{g2pf}
	\eeq
The reader who wishes to know more details of Picard-Fuchs equations 
associated with non-hyperelliptic curves may consult the work of 
Isidro. \cite{Isi} 

From these equations, we can construct differential equations satisfied by 
(\ref{sca}), but such differential equations are not helpful for a direct 
proof of scaling relation because they are partial differential equations. 
For this reason, we seek a more convenient form of Picard-Fuchs equations. 
A candidate is an ordinary differential form because the right hand sides 
of respective quantities in (\ref{sca}) are written in terms of only single 
variable derivative. 
Therefore, if Picard-Fuchs equations can take ordinary differential 
forms, it would be easy to obtain ordinary differential equations for $W$ 
and $w$. In addition to this, since they are ordinary differential 
equations, we can uniquely fix the basis of solution space and by this 
it becomes possible to verify (\ref{fo}). 

However, sadly, since the direct derivation of Picard-Fuchs equations in 
terms of single variable derivatives from the original period integral 
requires much labour, instead, let us try to derive such equations from 
(\ref{g2pf}). 

\begin{center}
\subsection{Ordinary differential form of G$_2$ Picard-Fuchs equations}
\end{center}

First, let us rewrite (\ref{g2pf}) in the form 
	\beq
	\left[\pa_{u}^2 -c_1 \pa_u \pa_v -c_2 \pa_v -c_3 \right]\lambda =0,
	\quad  
	\left[\pa_{v}^2 -d_1 \pa_u \pa_v -d_2 \pa_v 
	-d_3 \right]\lambda =0
	.\lab{re}
	\eeq
If there is any differential equation satisfied by $\lambda$, 
it must be a linear combination of 
the two equations in (\ref{g2pf}) and their differentiations. We would like 
to make an ordinary differential equation in terms of single moduli 
derivatives, e.g., $\pa_u \lambda , \pa_{u}^2 \lambda$, etc, 
but in order to obtain such equation from 
(\ref{re}) by repeating differentiations, mixed derivatives and other 
moduli derivatives like $\pa_u \pa_v\lambda$ or $\pa_v \lambda$ must be 
eliminated. These irrelevant derivatives can be dropped by representing 
them in terms of $\pa_u \lambda, \pa_u \pa_v \lambda$ and $\pa_v \lambda$. 

For example, regarding $\pa_{u}^2 \pa_v =\pa_v (\pa_{u}^2 )$, we get 
	\beq
	\pa_{u}^2 \pa_v \lambda =\left[\pa_v c_1 \pa_u \pa_v +
	\pa_v c_2 \pa_v +\pa_v c_3 +c_1 \pa_u (\pa_{v}^2 )
	+c_2 \pa_{v}^2 +c_3 \pa_v \right]\lambda
	\eeq
and further substituting $\pa_{v}^2 \lambda$ from (\ref{re}) into this 
expression, we can obtain 
	\beqa
	D\pa_{u}^2 \pa_v \lambda &=&\Bigl[
	c_1 d_3 \pa_u +(c_3 +c_2 d_2 +\pa_v c_2 +c_1 \pa_u d_2 )\pa_v \nm\\
	& & +(c_2 d_1 +c_1 d_2 +\pa_v c_1 +c_1 \pa_u d_1 )\pa_u \pa_v 
	+(c_2 d_3 +\pa_v c_3 +c_1 \pa_u d_3 )\Bigr] \lambda
	,\eeqa
where
	\beq
	D =1-c_1 d_1
	.\eeq	
In a similar manner, we can arrive at 
	\beqa
	D \pa_u \pa_{v}^2 \lambda &=&\Bigl[
	d_3 \pa_u +(c_3 d_1 +c_2 d_1 d_2 +d_1 \pa_v c_2 +\pa_u d_2 )
	\pa_v \nm\\
	& & +(c_2 d_{1}^2 +d_2 +\pa_u d_1 +d_1 \pa_v c_1 )\pa_u \pa_v 
	+(c_2 d_1 d_3 +\pa_u d_3 +d_1 \pa_v c_3 )\Bigr] \lambda
	.\eeqa
With these in remind, eliminating 
$\pa_u \pa_v \lambda, \pa_{u}^2 \pa_v \lambda$ and 
$\pa_u \pa_{v}^2 \lambda$, we can obtain the fourth-order ordinary 
differential equation satisfied by $\lambda$
	\beq
	\left[
	\pa_{u}^4 -\frac{1}{\Del}\left(\widetilde{c}_3\pa_{u}^3 +
	\widetilde{c}_2\pa_{u}^2 +\widetilde{c}_1 \pa_{u}+
	\widetilde{c}_0\right)\right]\lambda =0
	,\eeq 
where we have denoted only the equation associated with $u$-derivatives and 
the coefficients are given by 
	\beqa
	\Del&=&(-16u^6 v^2 +216u^3 v^3 -729v^4 +1024u^8 \mu -14976u^5 v\mu + 
    	54432u^2 v^2 \mu +421632u^4{\mu }^2\nm\\
	& &-2985984uv{\mu }^2 +47775744{\mu }^3)  
  	(648u^5 v^5 -8505u^2 v^6 -86016u^7 v^3\mu +1145664u^4 v^4\mu \nm\\
	& &-326592uv^5\mu +2097152u^9 v{\mu}^2 -24625152u^6 v^2{\mu}^2 - 
    	69672960 u^3 v^3 {\mu }^2 - 24634368 v^4 {\mu }^2 \nm\\
	& &-163577856 u^8 {\mu }^3 + 
    	2601123840 u^5 v {\mu }^3 +
	4824354816 u^2 v^2 {\mu }^3 -72704065536 u^4 {\mu }^4 \nm\\
	& &-88098471936 u v {\mu }^4 + 1091580198912 {\mu }^5),\nm\\
	-\widetilde{c}_0 &=&4( -648 u^7 v^7 + 18711 u^4 v^8 + 
    120192 u^9 v^5 \mu  - 3530304 u^6 v^6 \mu  + 
    1191186 u^3 v^7 \mu  + 2480058 v^8 \mu  \nm\\
	& &-11730944 u^{11} v^3 {\mu }^2 + 
    	291824640 u^8 v^4 {\mu }^2 -158824800 u^5 v^5 {\mu }^2 - 
    	356682204 u^2 v^6 {\mu }^2 \nm\\
	& &+100663296 u^{13} v {\mu }^3 -2121007104 u^{10} v^2 {\mu }^3 - 
    	42455384064 u^7 v^3 {\mu }^3 +51853167360 u^4 v^4 {\mu }^3 \nm\\
	& &+53651973888 u v^5 {\mu }^3 -13086228480 u^{12} {\mu }^4 + 
    	348024471552 u^9 v {\mu }^4 +3595729895424 u^6 v^2 {\mu }^4 \nm\\
	& &-4366362599424 u^3 v^3 {\mu }^4 -665208557568 v^4 {\mu }^4 - 
    	13420581617664 u^8 {\mu }^5 \nm\\
	& &-115104252690432 u^5 v {\mu }^5 + 
    	106830134820864 u^2 v^2 {\mu }^5 +1876392727805952 u^4 {\mu }^6 \nm\\
	& &-2191620144365568 u v {\mu }^6 +16936958366318592 {\mu }^7 ),\nm\\
	-\widetilde{c}_1 &=&4(-1944u^8 v^7 +44469u^5 v^8 +376320u^{10}v^5 
	\mu  -8745408u^7 v^6 \mu +3796632u^4 v^7\mu  -2480058uv^8 \mu \nm\\
	& &-25427968 u^{12} v^3 {\mu }^2 +563576832 u^9 v^4 {\mu }^2 - 
    206437248 u^6 v^5 {\mu }^2 + 
    1138989600 u^3 v^6 {\mu }^2 \nm\\
	& &-192735936 v^7 {\mu }^2 +402653184 u^{14} v {\mu }^3 - 
    7795113984 u^{11} v^2 {\mu }^3 - 
    41644523520 u^8 v^3 {\mu }^3 \nm\\
	& &-30988541952 u^5 v^4 {\mu }^3 +16587887616 u^2 v^5 {\mu }^3 - 
    41875931136 u^{13} {\mu }^4 +1020046344192 u^{10} v {\mu }^4 \nm\\
	& &+2862297907200 u^7 v^2 {\mu }^4 -3324905127936 u^4 v^3 {\mu }^4 - 
    69173305344 u v^4 {\mu }^4 \nm\\
	& &-33549868597248 u^9 {\mu }^5 
	-38241489125376 u^6 v {\mu }^5 +138188377423872 u^3 v^2 {\mu }^5 
	\nm\\
	& &+232190115840 v^3 {\mu }^5 +838333592764416 u^5 {\mu }^6 
	-3213346090254336 u^2 v {\mu }^6 \nm\\
	& &+24521257588359168 u {\mu }^7 ),\nm\\
	-\widetilde{c}_2 &=&2(-47952u^9 v^7 +1093014u^6 v^8 -1594323u^3 v^9 
	+6200145v^{10}+9277440u^{11}v^5 \mu \nm\\
	& &-215229312u^8 v^6 \mu +411156000u^5 v^7\mu -1285614828u^2 v^8\mu 	     
	-637534208 u^{13} v^3 {\mu }^2 \nm\\
	& &+14181875712u^{10}v^4{\mu}^2 -23978446848u^7 v^5{\mu}^2 + 
    	62058871872u^4 v^6{\mu}^2 +115596211968uv^7 {\mu}^2 \nm\\
	& &+9663676416u^{15}v{\mu}^3 -180703199232u^{12}v^2 {\mu}^3 - 
    	949095235584 u^9 v^3 {\mu}^3 \nm\\
	& &+1489634758656u^6 v^4 {\mu}^3 \nm\\
	& &-7237021925376 u^3 v^5 {\mu }^3 -4614778552320 v^6 {\mu }^3 - 
    	963146416128 u^{14} {\mu }^4 \nm\\
	& &+23697146511360 u^{11} v {\mu }^4 
	+48477813866496 u^8 v^2 {\mu }^4 -75294381441024u^5 v^3{\mu}^4 \nm\\
	& &+245288540749824 u^2 v^4 {\mu }^4 - 
    	763297561313280 u^{10} {\mu }^5 - 
    	174881136181248 u^7 v {\mu }^5 \nm\\
	& &-4532971954765824 u^4 v^2 {\mu }^5 - 
    	2597743016017920 u v^3 {\mu }^5 + 
    	14032107516985344 u^6 {\mu }^6 \nm\\
	& &+301566403503194112 u^3 v {\mu }^6 
	+88771668096319488v^2{\mu }^6 -5754950530929524736u^2{\mu }^7 ),
	\nm\\
	-\widetilde{c}_3 &=&2(-36288u^{10}v^7 +750384u^7 v^8 - 
    	2493180u^4 v^9 -6200145uv^{10}+7090176 u^{12} v^5 \mu \nm\\
	 & &-149257728u^9 v^6\mu +548581248u^6 v^7\mu +1031074272u^3 v^8\mu 
	 -119042784v^9 \mu \nm\\
	 & &-446693376 u^{14} v^3 {\mu }^2 + 
	 9473753088 u^{11} v^4 {\mu }^2 -34651597824 u^8 v^5 {\mu }^2 - 
    	59910223104 u^5 v^6 {\mu }^2 \nm\\
	& & -118816035840 u^2 v^7{\mu}^2 +7516192768 u^{16} v {\mu }^3 - 
    	136549761024u^{13}v^2{\mu}^3 -140014780416 u^{10} v^3 {\mu }^3 \nm\\
	 & &+4532540571648 u^7 v^4 {\mu }^3 + 
    	9634653609984u^4 v^5 {\mu }^3 +1729090768896uv^6 {\mu}^3 \nm\\
	& &-670014898176u^{15}{\mu}^4 +16657158242304u^{12}v{\mu}^4 - 
    28809835315200 u^9 v^2 {\mu }^4 \nm\\
	& &- 
    573860068982784 u^6 v^3 {\mu }^4 - 
    146014044291072 u^3 v^4 {\mu }^4 + 
    49247523569664 v^5 {\mu }^4 \nm\\
	& &-446693778653184u^{11}{\mu}^5 +2318736493117440u^8 v{\mu}^5 + 
    	27114473757081600 u^5 v^2 {\mu }^5 \nm\\
	& &-1492703816712192u^2 v^3{\mu}^5 -21106281041362944u^7{\mu}^6 - 
    	739515240667938816u^4 v{\mu }^6 \nm\\
	& &+19876959932645376uv^2{\mu}^6 +8787962215324975104 u^3 
	{\mu }^7 \nm\\
	& &-1154955987665289216 v{\mu }^7 ).
	\eeqa

Of course, a similar equation in terms of only $v$-derivatives can 
be found to follow by repeating the same algorithm (see Appendix). 
Below, we discuss only the case for $u$-derivatives. 

\begin{center}
\section{Differential equation for scaling relation} 
\end{center}

\begin{center}
\subsection{Ordinary differential equation for $W$} 
\end{center}

We can now construct an ordinary differential equation satisfied 
by $W$. To see this, let us define 
	\beq
	W_{ij}=\sum_{k=1}^2 (\pa_{u}^i a_k \pa_{u}^j a_{D_k}-\pa_{u}^j a_k 
	\pa_{u}^i a_{D_k})
	.\lab{ex}
	\eeq
Similar quantities are often used for a calculation of Yukawa couplings in 
the context of mirror symmetry. Notice that $W$ itself is given by 
$W=W_{01}$. 
From (\ref{ex}), we can derive some relations among various $W_{ij}$ 
	\[
	W^{\,'}=W_{02},\quad W^{\,''}=W_{12}+W_{03},\quad 
	W_{12}^{\,'}=W_{13}, \quad 
	W^{\,'''}=C_1 W+C_2 W^{\,'} +C_3 W_{03}+2 W_{13},\]
	\vspace{-1.5cm}
	\[
	W_{03}^{\,'}-C_3 W_{03}=C_1 W+C_2 W^{\,'}+W_{13},\quad 
	W_{13}^{\,'}-C_3 W_{13}=-C_0 W+C_2 W_{12}+W_{23},\]
	\vspace{-1.5cm}
	\beq
	W_{23}^{\,'}-C_3 W_{23}=-C_0 W^{\,'}-C_1 W_{12} ,\quad 
	C_i =\frac{\widetilde{c}_i}{\Del}
	\lab{rela}
	,\eeq
where $'=\pa/\pa u$. 

From (\ref{rela}), we can construct a sixth-order ordinary 
differential equation satisfied by $W$  
	\beq
	\left[\pa_{u}^5 +\widetilde{C}_4 \pa_{u}^4 +\widetilde{C}_3 
	\pa_{u}^3 +\widetilde{C}_2 \pa_{u}^2 +\widetilde{C}_1 \pa_u 
	+\widetilde{C}_0 \right]\pa_u W =0
	,\lab{33}
	\eeq
where $\widetilde{C}_i$ are very complicated and extremely lengthy rational 
functions in moduli and the scaling parameter $\mu$.

\begin{center}
\subsection{Basis of solutions} 
\end{center}

In order to obtain the basis of solutions to (\ref{33}), especially, 
to determine the indicial indices, by taking a Frobenius algorithm at 
the weak coupling region, it would be sufficient to consider (\ref{33}) 
with $\mu =0$. This is because the weak coupling solutions can be 
represented also in a series of $\mu$ and the only the lowest order terms 
of this series are relevant in the determination of indicial indices. 
Then (\ref{33}) turns to 
	\[
	\Bigl[u^5 (4u^3 -27v)^4 (608u^6 +32694u^3v -192465v^2 )
	\pa_{u}^5 \]
	\vspace{-1.5cm}
	\[+ 
	9u^4( 4u^3 -27v)^3 (7296u^9 + 463280u^6v - 1645836u^3v^2 - 
     	5196555v^3 )\pa_{u}^4 \]
	\vspace{-1.5cm}
	\[
	+2u^3 (4u^3 -27v)^2
   	(1162496u^{12}+82291704u^9v -292007736u^6v^2 -238968387u^3v^3 \]
	\vspace{-1.5cm}
	\[
	-2806139700v^4 )\pa_{u}^3 +12u^2 (4u^3 -27v)(2687360u^{15} 
	+205589136u^{12}v\]
	\vspace{-1.5cm}
	\[
	-1225216692u^9v^2 +1404464913u^6v^3 +6080288652u^3v^4 -
	37882885950v^5 )\pa_{u}^2 \]
	\vspace{-1.5cm}
	\[
	 +8u( 20422720u^{18}+1666413804u^{15}v-16557855750u^{12}v^2 \]
	\vspace{-1.5cm}
	\[
	+33777521721u^9v^3 +1420502427u^6v^4 +879666475221u^3v^5 - 
     	3068513761950v^6 )\pa_u \]
	\vspace{-1.5cm}
	\[
	+40(1337600u^{18}+125474352u^{15}v - 
     	812719908u^{12}v^2 +1338655410u^9 v^3 +7591437855u^6 v^4 \]
	\vspace{-1.5cm}
	\beq
	-162446987646u^3 v^5 +613702752390v^6 )\Bigr]\pa_u W=0
	,\lab{222}
	\eeq
and therefore we get the following set of indicial indices 
	\beq
	\nu =(\nu_1 ,\nu_2 ,\nu_3 ,\nu_4 ,\nu_5 )=(1,2,3,5,8)
	\eeq
for 
	\beq
	W^{\,'}=u^{\nu}\widetilde{W}
	.\eeq

According to Frobenius algorithm, in general, there are two types of 
solutions due to the difference of two indices, one of which is a 
regular series and the other is logarithmic. In fact, the reader may be 
familiar to the following well-known fact: 
	\begin{itemize}
	\item
	Suppose that $\nu_1$ and $\nu_2$ are the indicial indices to a 
	second-order linear ordinary differential equation (unknown $y$ and 
	variable $x$) with regular singularities, say, at $x=0$. 
	Then the general solution is given in the form 
	\begin{enumerate}
	\item
	$\displaystyle 
	y= c_1 x^{\nu_1}\sum_{k=0}^{\infty}A_k x^k +c_2  x^{\nu_2}
	\sum_{k=0}^{\infty}B_k x^k ,\quad ( \nu_1 -\nu_2 \neq \mbox{ integer})
	$
	\item
	$\displaystyle 
	y=c_1 x^{\nu_1}\sum_{k=0}^{\infty}A_k x^k +c_2 \left[ c x^{\nu_1} 
	\sum_{k=0}^{\infty}A_k x^k \ln x+ x^{\nu_2}\sum_{k=1}^{\infty}B_k x^k \right]
	,\quad ( \nu_1 -\nu_2 =\mbox{ integer}),$
	\end{enumerate}
	where $c_i$ are integration constants, $c$ is a constant to be determined 
	according to the difference of the indices (typically, 
	$c=1$ when $\nu_1 =\nu_2 $), $A_k$ and $B_k$ are 
	independent of $x$. In both cases, when $\nu_1 \neq \nu_2 $, we can 
	assume $\nu_2 <\nu_1 $ without loss of generality. 
	\end{itemize}
Summarizing this, we can say that in any case the solution 
can be factored by $x^{\nu_2}$ associated with the smaller index. 
Similar one to this fact also holds for a higher rank ordinary differential 
equation (of course, in this case, the solution may involve power of 
logarithm due to the difference of indicial indices). 
Therefore, for (\ref{33}), $W^{\,'}$ takes the form 
	\beq
	W^{\,'} =u^{\nu_1}\sum_{i=1}^{5}\rho_i u^{\nu_i -\nu_1}f_i (u,v,\mu )
	,\eeq
where $\rho_i$ are some constants and $f_i$ are functions 
whose lowest order in expansion is logarithm or function 
which is independent of $u$. Integrating this gives the following function form 
	\beq
	W=c(v)+uf
	,\lab{wwww}
	\eeq
where $c(v)$ is a function which may depend on $v$, by using a function $f$ 
whose lowest order in expansion is logarithm or function independent of $u$. 

In order to make a contact with the weak coupling behavior, we must impose 
some ``initial condition''. For this purpose, let us recall the definition 
of $W$ in (\ref{sca}). Substituting the solutions to Picard-Fuchs equations 
into (\ref{sca}), we will be able to compare it with the right hand side of 
(\ref{wwww}). Of course, as a matter of fact, since the function form of 
$W$ is now uniquely determined as in (\ref{wwww}), it is not necessary to 
know the solutions 
to Picard-Fuchs equations at all order in $\mu$ and is enough to know them 
only at the lowest order level. In fact, proceeding in this manner with the 
aid of the weak coupling behavior of periods from (\ref{lukeskywalker}) (or 
explicitly solving (\ref{222}) or using Ito's result \cite{Ito}), 
we can see that the second term of (\ref{wwww}) is suppressed and 
$c(v)=i/(4\pi )$. This indicates that 
	\beq
	\sum_{i=1}^2 (a_i \pa_u a_{D_i} -a_{D_i}\pa_u a_i )=
	\frac{i}{4\pi}
	\lab{38}
	\eeq
holds exactly! Therefore, the scaling relation found by Ito \cite{Ito} 
in the G$_2$ gauge theory based on the spectral curve (\ref{12}) is 
actually an exact expression. 

We can conclude that the scaling relation
	\beq
	\sum_{i=1}^2 (a_i \pa_v a_{D_i} -a_{D_i}\pa_v a_i )=0
	\lab{39}
	\eeq
holds exactly by repeating a similar discussion. 

\begin{center}
\section{Summary}
\end{center}

In this paper, we have proved the scaling relation of prepotential of G$_2$ 
Yang-Mills theory by using ordinary differential form of Picard-Fuchs 
equations. The direct verification of the scaling relation (when multiple 
moduli are included) is very complicated, but it would be needless to say 
that our direct method presented here can be applied for any gauge group 
cases. That is, when one wish to establish the scaling relations like 
(\ref{38}) and (\ref{39}): 
	\begin{enumerate}
	\item
	Represent the Picard-Fuchs equations in terms of single moduli 
	derivatives. 
	
	\item
	Consider the ordinary differential equation satisfied by 
	scaling relation using the above Picard-Fuchs equations. 

	\item
	Fix the basis of solutions to this equation. 

	\item
	Compare the result with that from solutions to Picard-Fuchs 
	equation. 

	\end{enumerate}
Note that a direct derivation of (\ref{fo}) by Picard-Fuchs equations as a 
system of partial differential equations is more involved than our 
presentation here. 

In contrast with the classical gauge group cases, there are many problems 
concerning exceptional gauge group cases and in order to get more insight 
into these exceptional gauge theories application of Whitham 
hierarchy \cite{GMMM}  is 
necessary. To to this is very hard because of the complicated 
singularity structure of the spectral curves, but, also the problem of 
scaling relation in these cases should be understood in this framework.

\begin{center}
\section*{Appendix: Another scaling relation}
\end{center}

\renewcommand{\theequation}{A\arabic{equation}}\setcounter{equation}{0}

In this appendix, we show that the Picard-Fuchs equation in terms 
of only $v$-derivatives and the ordinary differential equation for 
$w$. 

The Picard-Fuchs equation can be represented as the fourth-order 
equation and takes the form
	\[ \Biggl[
	-9(37u^8 -126u^5 v+81u^2 v^2 -81216u^4 \mu +171072uv\mu + 
     	36578304{\mu}^2 ) \]
	\vspace{-1.3cm}
	\[+24(68u^{11} -1422u^8 v+12393u^5 v^2 -37179u^2 v^3 +61776u^7 \mu - 
	2220048u^4 v\mu +10707552uv^2 \mu \]
	\vspace{-1.3cm}
	\[ +204166656u^3{\mu}^2 -1169012736v
	{\mu }^2 )\pa_v -24( 24 u^{14} - 1004 u^{11} v + 
     15885 u^8 v^2 - 108540 u^5 v^3 \]
	\vspace{-1.3cm}
	\[ +273375u^2 v^4 +45648u^{10}\mu -1823040u^7 v\mu +19801584u^4 v^2
	 \mu -70123968uv^3 \mu +40061952u^6{\mu}^2 \]
	\vspace{-1.3cm}
	\[ -724847616u^3 v{\mu}^2 +4938071040v^2 {\mu}^2 -32356122624u^2 
	{\mu}^3 )\pa_{v}^2 \]
	\vspace{-1.3cm}
	\[ -32(u^4 -9uv+864\mu )(32u^{10}v-792u^7 v^2 +6804u^4 v^3 - 
	19683uv^4 +5760u^9 \mu -81216u^6 v\mu -69984u^3 v^2 \mu \]
	\vspace{-1.3cm}
	\[ +2519424v^3 \mu +12130560u^5 {\mu}^2 -94058496u^2 v{\mu}^2 +
	2149908480u{\mu}^3)\pa_{v}^3 \]
	\vspace{-1.3cm}
	\[ +16(u^4 -9uv+864\mu )^2 (-16u^6 v^2 +216u^3 v^3 -729v^4 + 
     1024u^8 \mu -14976u^5 v\mu +54432u^2 v^2 \mu +421632u^4{\mu}^2 \]
	\vspace{-1.3cm}
	\beq -2985984uv{\mu }^2 +47775744{\mu}^3 )\pa_{v}^4 \Biggr]
	\lambda =0 
	.\eeq

Then the differential equation for $w$ with $\mu=0$ is given by 
	\[\Biggl[ 
	-6 ( 68u^{21}-1458u^{18}v +107406 u^{15} v^2 - 2216889 u^{12} v^3 + 
     	21611934u^9 v^4 -121168548u^6 v^5 \]
	\vspace{-1.3cm}
	\[ +382637520 u^3 v^6 - 516560652v^7 )+ 
  6 ( 24 u^{24} - 5656 u^{21} v + 
     298890 u^{18} v^2 - 7376508 u^{15} v^3 \]
	\vspace{-1.3cm}
	\[ +104622435u^{12}v^4 -906638346u^9 v^5 +4757459832u^6 v^6 -
	13937571666u^3 v^7 +17563062168v^8 )\pa_v \]
	\vspace{-1.3cm}
	\[ +27(4u^3 -27v)(u^3 -9v)v(32u^{18} -2216 u^{15} v + 
     61950 u^{12} v^2 - 883845 u^9 v^3 + 
     6896583 u^6 v^4 \]
	\vspace{-1.3cm}
	\[ -28199178u^3 v^5 +47475396v^6 )\pa_{v}^2 +( 4 u^3 - 27 v )^2 
   ( u^3 - 9 v )^2 v^2 
   ( 392 u^{12} - 15108 u^9 v + 226503 u^6 v^2 \]
	\vspace{-1.3cm}
	\[ -1515348u^3 v^3 +3831624v^4 )\pa_{v}^3 +5(4u^3 -27v)^3 (u^3 -9v)^3 v^3 
   ( 8 u^6 - 144 u^3 v + 729 v^2 )\pa_{v}^{4} \]
	\vspace{-1.3cm}
	\beq + ( 4 u^3 - 27 v )^4 ( u^3 - 9 v )^4 v^4 \pa_{v}^5 \Biggr]W=0
	.\lab{er}	
	\eeq
The set of indicial indices for (\ref{er}) is found to be 
	\beq
	\nu =(1,1/2,1/2,-1/2,-1/2), \quad w=v^{\nu}\widetilde{w}
	.\eeq

\begin{center}

\end{center}
\end{document}